\documentclass[aps,groupedaddress,superscriptaddress,amssymb,amsmath,amsbsy,amsfonts,twocolumn,pra]{revtex4}

\usepackage{amssymb,amsmath,amsthm,color,graphicx,times,graphicx}
\usepackage{hyperref}
\usepackage{enumerate}
\usepackage{subfigure}
\usepackage{dsfont}
\usepackage{times}
\usepackage{txfonts}
\usepackage{bbold}

\usepackage{extarrows}

\newcommand*{\cl}[1]{{\mathcal{#1}}}
\newcommand*{\bb}[1]{{\mathbb{#1}}}
\newcommand{\ket}[1]{\left|#1\right>}

\newcommand*{\tn}[1]{{\textnormal{#1}}}
\newcommand*{\1}{{\mathbb{1}}}
\newcommand{\T}{\mbox{$\textnormal{Tr}$}}

\theoremstyle{plain}

\theoremstyle{definition}

\date{\today}

\begin{document}
	
\title{High-dimensional private quantum channels and regular polytopes}
	
\author{Junseo Lee}
\email{junseo0218@yonsei.ac.kr}
\affiliation{School of Electrical and Electronic Engineering, Yonsei University, Seoul 03722, Korea}
\author{Kabgyun Jeong}
\email{kgjeong6@snu.ac.kr}
\affiliation{Research Institute of Mathematics, Seoul National University, Seoul 08826, Korea}
\affiliation{School of Computational Sciences, Korea Institute for Advanced Study, Seoul 02455, Korea}
	
\date{\today}
\pacs{
03.67.Hk, 
03.67.Ac  
}
	
\begin{abstract}
As the quantum analog of the classical one-time pad, the private quantum channel (PQC) plays a fundamental role in the construction of the maximally mixed state (from any input quantum state), which is very useful for studying secure quantum communications and quantum channel capacity problems. However, the undoubted existence of a relation between the geometric shape of regular polytopes and private quantum channels in the higher dimension has not yet been reported. Recently, it was shown that a one-to-one correspondence exists between single-qubit PQCs and three-dimensional regular polytopes (i.e., regular polyhedra). In this paper, we highlight these connections by exploiting two strategies known as a generalized Gell-Mann matrix and modified quantum Fourier transform. More precisely, we explore the explicit relationship between PQCs over a qutrit system (i.e., a three-level quantum state) and regular 4-polytope. Finally, we attempt to devise a formula for connections on higher dimensional cases. 
\end{abstract}
	
\maketitle

\section{Introduction}
Modern cryptographic systems essentially rely on the computational-complexity assumption for their security, whereas a quantum communication primitive known as a private quantum channel (PQC) achieves its safety under information-theoretic conditions. The PQC, first proposed by Ambainis \emph{et al.}~\cite{AMTW00}, provides a fundamental and perfectly \emph{secure} way to transmit a quantum state from a sender, Alice, to a receiver, Bob, by using pre-shared classical secret keys generated by a quantum key distribution (QKD) scheme. As a kind of completely positive and trace preserving map (i.e., quantum channels)~\cite{NC00,W13}, the PQC transforms any quantum state into a maximally mixed state (MMS) in a given Hilbert space.
	
Because the output of PQCs always fulfills the genuine maximally mixed state, which is a quantum state with a maximal von Neumann entropy (i.e., strong against any type of attack), they can be used to construct a secure quantum network or quantum internet~\cite{K08,PWE+15,VPZ+16,WEH18} with the help of quantum teleportation~\cite{BBC+93} as well as QKD protocols~\cite{BB84,E91,B92} for emerging quantum communication technologies.  Another main feature of the PQC at the purely theoretical level is related to a phenomenon known as \emph{superadditivity} on quantum channel capacity problems~\cite{SY08,H09,LWZG09}. In particular, a PQC and its dual (i.e., its complementary PQC) reportedly form a subtle counter-example to the additivity, especially on the classical capacity~\cite{H09} on quantum channels (the PQCs), owing to quantum entanglement~\cite{HHHH09}.
	
In traditional geometry, it is well known that an infinite number of regular polygons exists in a two-dimensional plane and five regular polyhedra exist in three-dimensional space~\cite{C73}. Few recent efforts to relate the regular polytopes to quantum information theory, for example, a construction on Bell's inequalities~\cite{TG20,BK20} have been reported. Classifying or proving the existence of a higher dimensional ($d>3$) regular $d$-polytope is not a trivial problem. However, the regular 4-polytope (with a cell) was well classified by several mathematicians many years ago. 
	
Here, we attempt to devise an approach to connect the structure of PQCs to higher dimensional regular polytopes, under the constraint of preserving the maximal output entropy. To this end, we need to exploit the notion of an isotropic (or unitarily invariant) measure on the unitary group, and two modifications of the Gell-Mann matrix~\cite{G62} and quantum Fourier transform~\cite{C02}. In this study, we highlight a new connection between qutrit-based PQCs and the regular 4-polytope by generalizing our previous research on qubit PQCs and regular polyhedra~\cite{JLC+18} equipped with an isotopic measure.   
	
The remainder of this paper is organized as follows. In Sec.~\ref{pre}, we describe the basic concept of the PQC, and provide definitions for several relevant and meaningful quantities, such as the isotropic measure and the regular polytopes, especially for the convex regular 4-polytope. To relate the PQCs and polytopes, we introduce a new notion of a hypervector in Sec.~\ref{hypervec}. In Sec.~\ref{main}, we derive our main results (using two methods) on the relationship between qutrit-based PQCs and the regular 4-polytope, which is a generalization of qubit-based PQCs. Especially, in Sec.~\ref{qutrit}, we briefly argue a universal strategy for the connection over higher dimensional cases. Finally, discussions and remarks are presented in Sec.~\ref{conclusion}, and a few intriguing questions are raised for future work.
	
\section{Preliminaries} \label{pre}
\subsection{Concept of a private quantum channel} \label{definitions}
Here, we briefly review the mathematical definition and related results of private quantum channels. Before providing the details, we explain our notations. Let $\cl{B}(\bb{C}^d)$ denote the set of linear operators from the Hilbert space $\bb{C}^d$ to itself, and let $\tn{U}(d)\subset\cl{B}(\bb{C}^d)$ be the unit group on $\bb{C}^d$. Let us define a quantum channel as $\Lambda:\cl{B}(\bb{C}^d)\to \cl{B}(\bb{C}^d)$, which is a linear, completely positive, and trace-preserving map. For any quantum state $\rho\in\cl{B}(\bb{C}^d)$, the quantum channel is conveniently denoted as $\Lambda:\rho\mapsto\Lambda(\rho)$ in $\cl{B}(\bb{C}^d)$.
	
Generally, we consider a quantum channel $\Lambda:\cl{B}(\bb{C}^d)\to \cl{B}(\bb{C}^d)$ to be a $\varepsilon$-PQC (or approximate PQC)~\cite{J12} if it satisfies that
\begin{equation} \label{eq:apqc}
\left\|\Lambda(\rho)-\frac{\1}{d}\right\|_p\le\frac{\varepsilon}{d^{\frac{p-1}{p}}},
\end{equation}
where $\varepsilon$ is a small (but non-negative) real number, and $\1$ denotes the $d\times d$ identity matrix. The Schatten $p$-norm or matrix norm, $\|\cdot\|_p$, is defined by $\|M\|_p=\sqrt[p]{\T(M^\dag M)^{p/2}}=\left(\sum_js_j^p(M)\right)^{1/p}$, where $s_j(M)$ denotes the singular values of any matrix $M$. If the parameter $\varepsilon=0$, that is, $\Lambda(\rho)=\frac{\1}{d}$, we say that the map $\Lambda(\cdot)$ is a \emph{complete} PQC. To obtain Eq.~\eqref{eq:apqc}, we can straightforwardly combine the definitions over the operator norm~\cite{HLSW04} and the trace norm~\cite{DN06} by using McDiarmid's inequality.
	
\begin{figure}[t!]
\includegraphics[width=\linewidth]{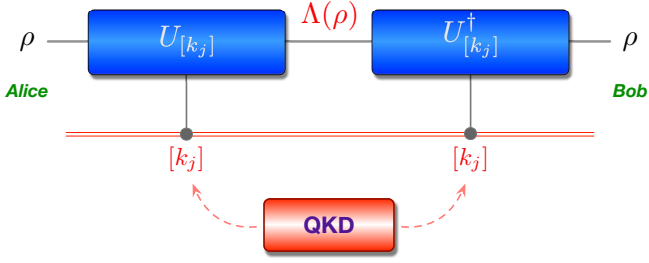}
\caption{Schematic diagram of a PQC. We assume that Alice and Bob shared a secret key $K=[k_j]$ via quantum key distribution (QKD) protocols. Should Alice wish to encode a quantum state $\rho$ through the PQC $\Lambda$, she applies unitary operations, $U_{[k_j]}$, on the input state depending on the key set $[k_j]$. Thus, $\Lambda(\rho)$ is equivalent to the maximally mixed state $\tfrac{\1}{d}$. At the end of the PQC, the receiver Bob can recover the original quantum state $\rho$ by exploiting the inverse units over $\Lambda(\rho)$.}
\label{Fig1}
\end{figure}
	
The advantage of $\varepsilon$-PQC is that it is possible to reduce its cardinality in terms of the unitary operations required, from the optimal case ($d^2$) to the approximate regimes of $\cl{O}(d\log d)$~\cite{HLSW04} or $\cl{O}(d)$~\cite{A09}. Here, we notice that the dependence of the optimality of PQCs on the input dimension of the quantum state was determined by several groups~\cite{BR03,NK06,BZ07}. However, in this paper, we only consider the optimal schemes for matching regular polytopes. In addition, several considerations are known to exist on continuous-variable PQCs~\cite{B05,JKL15,JL16,BSZ20} as well as on a sequential version~\cite{JK15}. However, the main purpose of this work is to contribute to the construction of (secure) key-dependent PQCs (but satisfying a maximal entropy condition) over a set of unitaries provided by the key set $K=[k_j]:=\{k_1,k_2,\ldots\}$ for all $j$. For a given key set $K$, we can create a private quantum channel in the form of 
\begin{equation} \label{eq:pqc}
\Lambda_K(\rho)=\frac{1}{|K|}\sum_{j=1}^{|K|}U_{k_j}\rho U_{k_j}^\dag,
\end{equation}
where $U_{k_j}$ are unitary operators induced from each component $k_j\in K$. For an example of the optimal case (i.e., $|K|=4$), the key set $K$ can be constructed from Pauli matrices, that is, $K=\{\1,X,Y,Z\}$ with $X=\begin{pmatrix} 0 & 1 \\ 1 & 0 \end{pmatrix},Y=\begin{pmatrix} 0 & -i \\ i & 0 \end{pmatrix},X=\begin{pmatrix} 1 & 0 \\ 0 & -1 \end{pmatrix}$. Then, for every $\rho\in\cl{B}(\bb{C}^2)$, we have
\begin{equation*}
\Lambda_K(\rho)=\frac{1}{4}(\1\rho\1^\dag+X\rho X^\dag+Y\rho Y^\dag+Z\rho Z^\dag)=\frac{\1}{2}.
\end{equation*}
Here, we can observe that the channel output $\Lambda_K(\rho)$ is the exact two-dimensional maximally mixed state. As mentioned above, key set $K$ can be obtained from QKD protocols. (See the scheme in Fig.~\ref{Fig1}). 
	
In our previous study~\cite{JLC+18}, we found a five key set $\{K_P=4,K_H=6,K_O=8,K_D=12,K_I=20\}$ corresponding to PQCs in terms of $\{\cl{N}_P,\cl{N}_H,\cl{N}_O,\cl{N}_D,\cl{N}_I\}$, where the subscripts denote Pauli (or tetrahedron), hexahedron, octahedron, dodecahedron, and icosahedron, respectively. The main requirement these constructions need to meet is that they always preserve the maximal von Neumann entropy because all outputs of each of the PQCs are exactly the maximally mixed state in the Hilbert space $\bb{C}^2$. These constructions are followed by the extension of the Pauli matrices via a complex rotation matrix of
\begin{equation} \label{2d}
R_\tn{c}(\theta)=\begin{pmatrix} \cos\theta & -i\sin\theta \\ i\sin\theta & \cos\theta \end{pmatrix},
\end{equation}
where the angle $\theta\in[0,2\pi]$ is a real number. The main objective of this research is to generalize the qubit-based PQCs above to a qutrit-based PQC as well as to find a formulation on the higher dimensional cases.
	
\subsection{Isotropic measure} \label{measure}
To intuitively obtain the relationship between PQCs and regular polytopes, we need to review the notion of an isotropic measure on the unitary group $\tn{U}(d)$. The isotropic measure for quantum states is formally defined as follows ~\cite{A09}: For any quantum state $\rho\in\cl{B}(\bb{C}^d)$, a probability measure $\mu$ on the unitary group $\tn{U}(d)$ is said to be \emph{isotropic}, if it holds that
\begin{equation} \label{eq:iso}
\int_{\tn{U}(d)}U\rho U^\dag\tn{d}\mu=\frac{\1}{d}.
\end{equation}
In addition, a random vector $\vec{v}$ generated by $U\in \tn{U}(d)$ is known to be isotropic if its law is isotropic. Conceptually, this implies that the integration over all random vectors (generated by $U$) equates to zero (i.e., the center of mass).
	
In the case of a discrete measure, the structure of PQC in Eq.~\eqref{eq:pqc} corresponds to that of the exact isotropic measure. As an example, the set of Pauli matrices $\{\1,X,Y,Z\}$ is isotropic and the set of corresponding random vectors, namely, $\{\vec{v}_{\1},\vec{v}_X,\vec{v}_Y,\vec{v}_Z\}$ is also isotropic. Thus, by definition, the sums of the actions of the Pauli matrices and random vectors are $\frac{\1}{2}$ and 0, respectively. We notice that the Haar measure on $\tn{U}(d)$ is also an isotropic measure.
	
In this study, we connect the private quantum channels with a key set $K$ to the regular polytopes beyond the low-dimensional cases through Eq.~\eqref{eq:iso}. However, not only is the extension quite complex, even in the case of four dimensions, but the higher dimensional polytopes are also not well defined. 
	
Before discussing the relationship, we briefly review the regular polytopes in four-dimensional (Euclidean) space.

\subsection{Regular 4-polytope} \label{polytope}
All the classifications and proofs of existence of the regular $d$-polytope are very difficult problems in geometry~\cite{C73}. Here, we only take into account the regular \emph{convex} 4-polytope as a natural matching for the three-level quantum state (i.e., qutrit) because the geometric shape (in terms of the Bloch sphere) of any quantum state satisfies the convex set and a unit sphere of a given dimension.
	
The regular convex 4-polytope was first introduced by Schl\"{a}fli. Six types of convex-type polytopes that are four-dimensional analogues of the three-dimensional regular polyhedra (i.e., Platonic solids) exist. The existence of the regular convex 4-polytope, which is generally denoted by a Schl\"{a}fli symbol $[\alpha,\beta,\gamma]$, is constrained by cells (i.e., three-dimensional regular polyhedra) and dihedral angles (see Table~\ref{table:pqc} below). In addition, each polytope in geometry can be classified by intrinsic symmetric groups as in Table~\ref{table:pqc}, and is generally known as a Coxeter group~\cite{C73,J18}. 
	
\begin{table*}
\caption{Summary of regular 4-polytopes versus qutrit-based PQCs.}
\begin{ruledtabular}
\begin{tabular}{l c c c c c c c} 
{} & {-} &  Simplex ($S$) & Hypercube ($H$) & Tesseract ($T_1$) & Octaplex ($O$) & Dodecaplex ($D$) & Tetraplex ($T_2$)  \\
\hline
{\bf Orthographics} & {Conj.}  & \begin{minipage}{0.6in}\includegraphics[width=\textwidth]{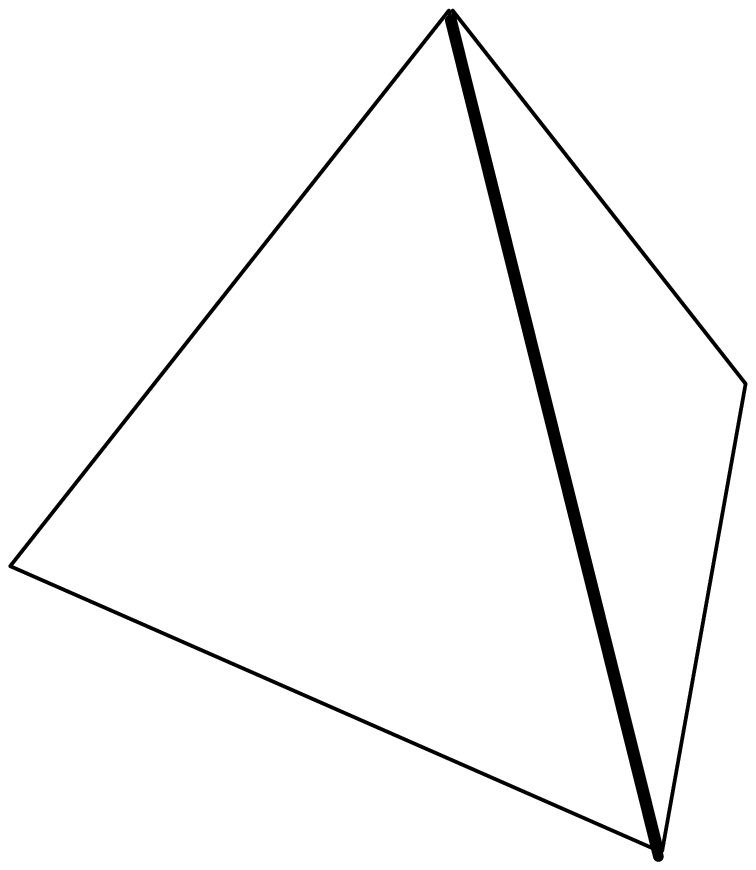}\end{minipage} & \begin{minipage}{0.6in}\includegraphics[width=\textwidth]{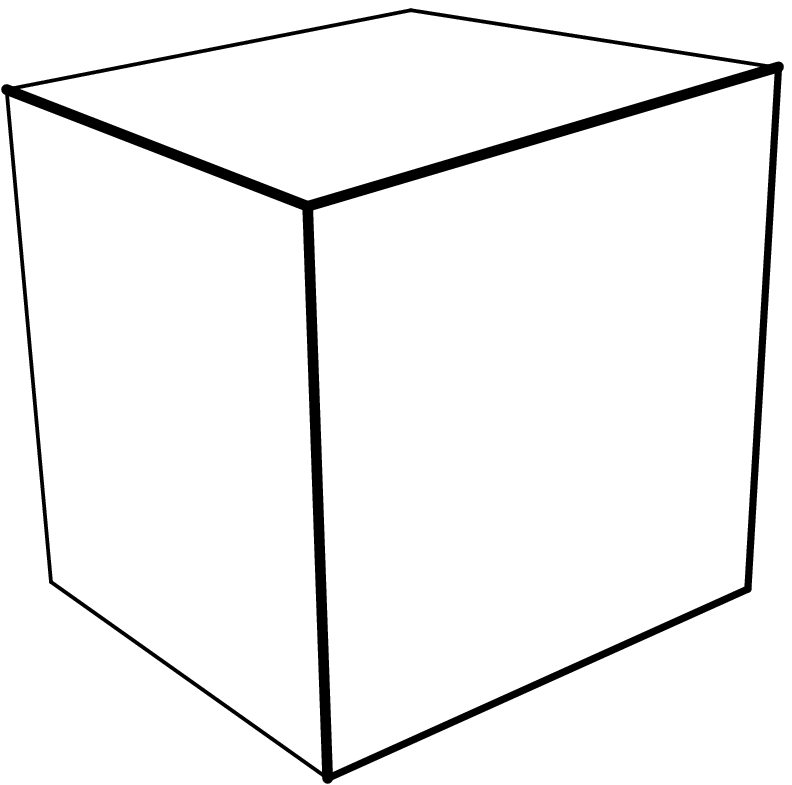}\end{minipage} & \begin{minipage}{0.6in}\includegraphics[width=\textwidth]{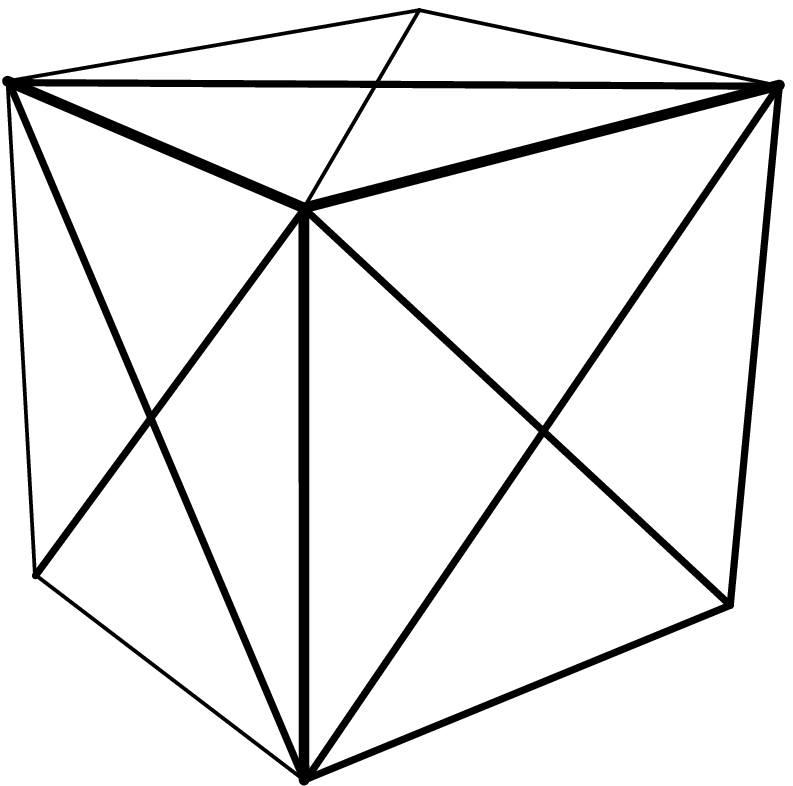}\end{minipage} & \begin{minipage}{0.57in}\includegraphics[width=\textwidth]{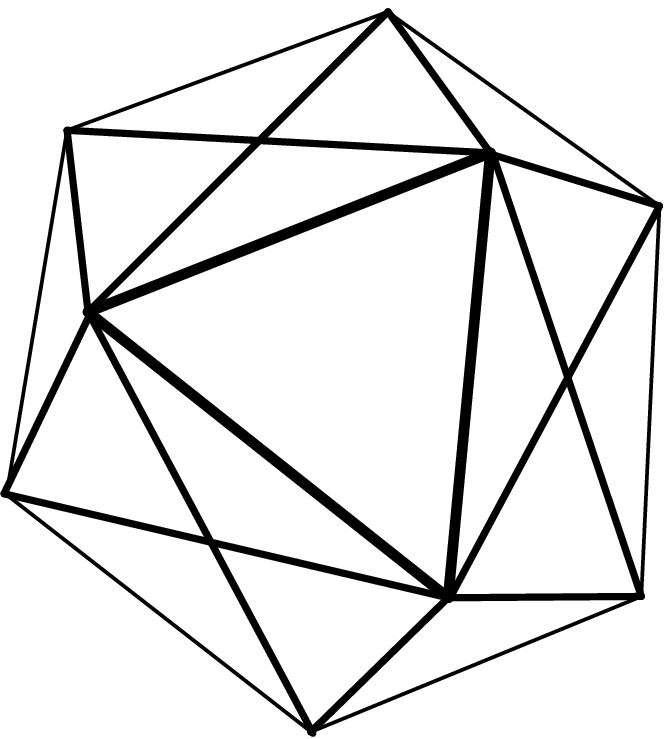}\end{minipage} & \begin{minipage}{0.6in}\includegraphics[width=\textwidth]{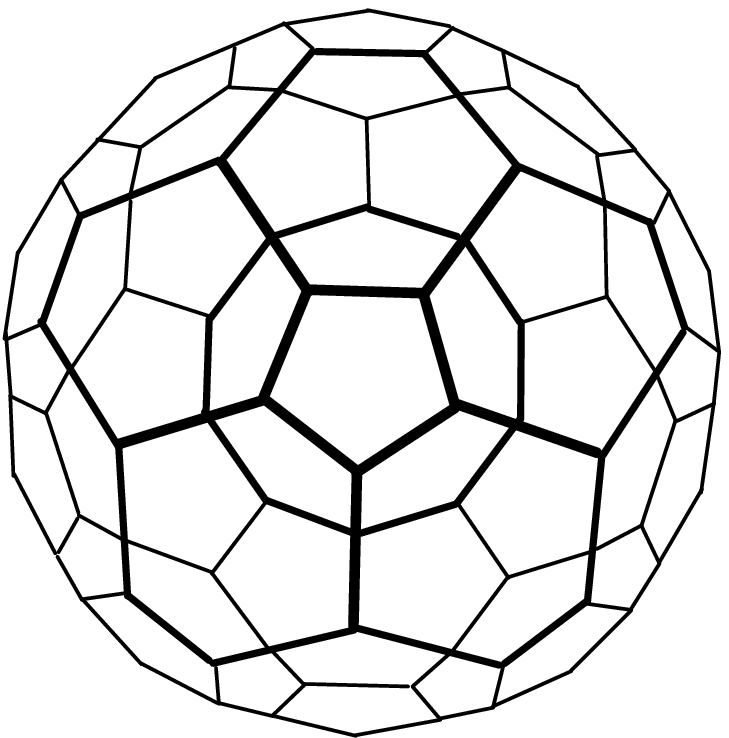}\end{minipage} & \begin{minipage}{0.6in}\includegraphics[width=\textwidth]{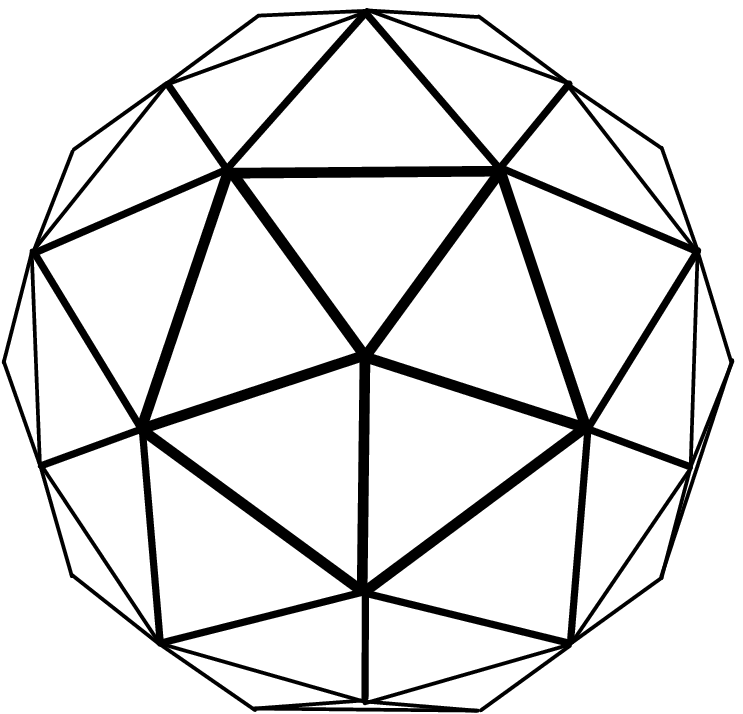}\end{minipage} \\
{\bf Cell} & - & 5 & 8 & 16 & 24 & 120 & 600 \\
{\bf Symmetric group} & {-}  & $A_4=[3,3,3]$ & $B_4=[4,3,3]$ & $B_4=[3,3,4]$ & $F_4=[3,4,3]$ & $H_4=[5,3,3]$ & $H_4=[3,3,5]$ \\
\hline
{\bf PQC $\Lambda$} & $\Lambda_L$ & $\Lambda_{S}$ & $\Lambda_{H}$ & $\Lambda_{T_1}$ & $\Lambda_{O}$ & $\Lambda_{D}$ & $\Lambda_{T_2}$ \\
\hline
{\bf Hypervector $\mathbf{V}$} & {-} & 5 & 8 & 16 & 24 & 120 & 600 \\
{\bf Basis vector $\vec{v}$} & {9} & 20 & 48 & 96 & 192 & 1440 & 12000 \\
{\bf Cardinality $|K|$} & 9 & 20 & 48 & 96 & 192 & 1440 & 12000 \\
{\bf Optimality} & $\bigcirc$ & $\times$ & $\times$ & $\times$ & $\times$ & $\times$ & $\times$ \\
{\bf Security} & $\bigcirc$ & $\bigcirc$ & $\bigcirc$ & $\bigcirc$ & $\bigcirc$ & $\bigcirc$ & $\bigcirc$
\end{tabular}
\end{ruledtabular}
\label{table:pqc}
\end{table*}

\section{Hypervector and regular convex 4-polytope} \label{hypervec}
Let $\mathbf{V}$ be a set of vectors, that is, $\mathbf{V}_t=(\vec{v}_1,\vec{v}_2,\ldots,\vec{v}_t)$, and we term it a \emph{hypervector}. It exactly corresponds to a $j$-component set of a regular polyhedron. We notice that a hypervector is a vector in four-dimensional space, but $\vec{v}_j$'s are three-dimensional. As an example, $\mathbf{V}_4=(\vec{v}_1,\vec{v}_2,\vec{v}_3,\vec{v}_4)$ can be interpreted as a schematic in the form of
\begin{equation} \label{hypervector}
\begin{minipage}{0.6in}\includegraphics[width=\textwidth]{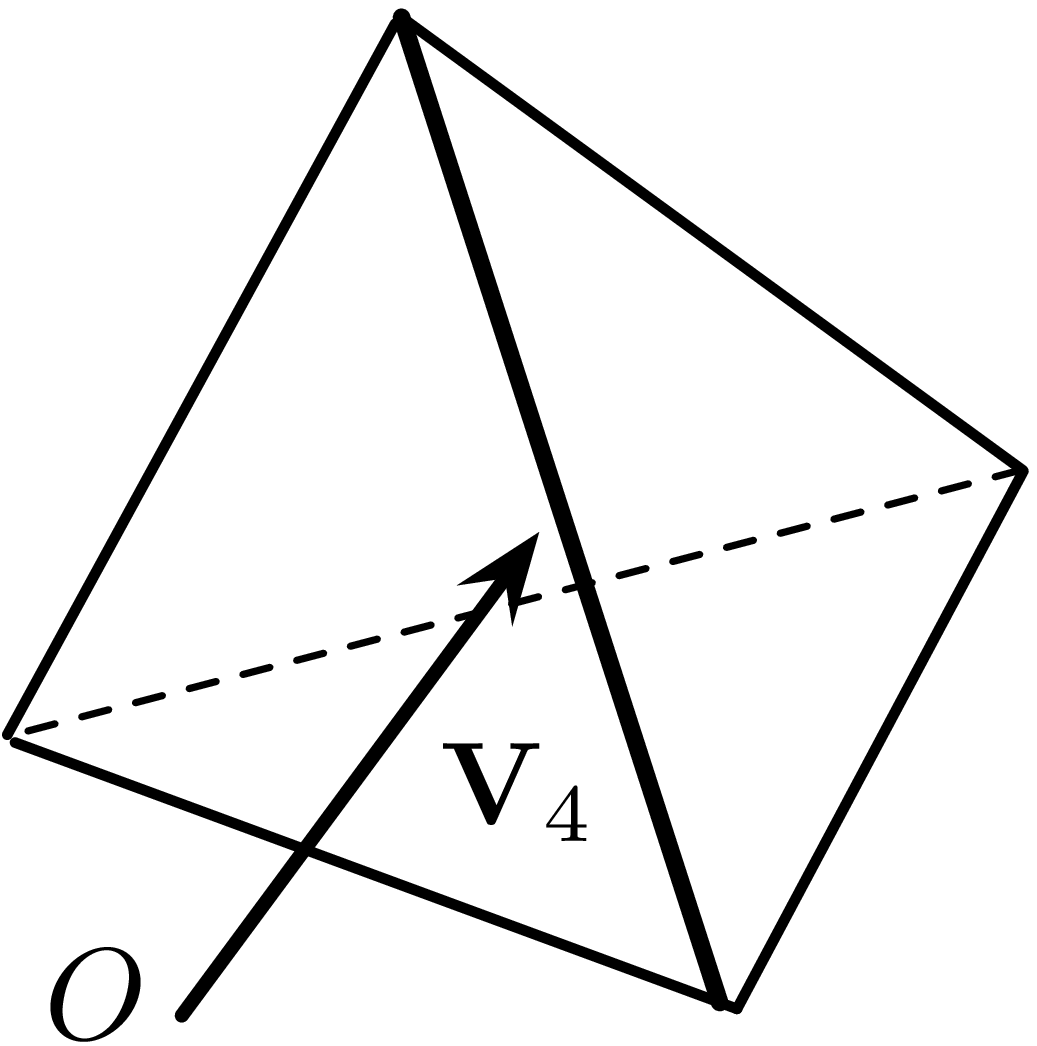}\end{minipage}\;\;\;\;  \equiv\;\;\;\;  \begin{minipage}{0.6in}\includegraphics[width=\textwidth]{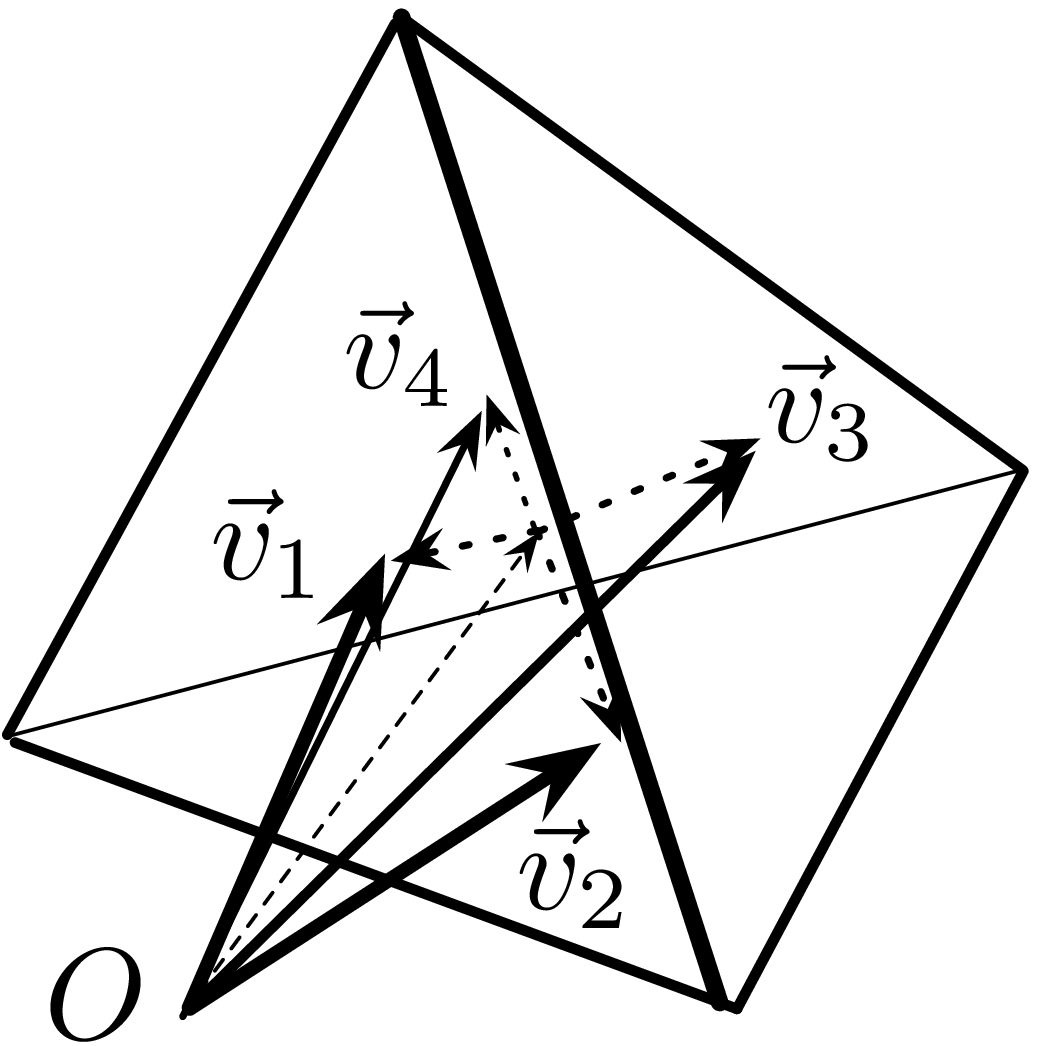}\end{minipage},
\end{equation}
where $\vec{v}_t$'s are four vectors in the tetrahedron (i.e., one of the regular 3-polytopes). Interestingly, our hypervector $\mathbf{V}_4$ corresponds with the Pauli matrices in a one-to-one manner. In this context, we can naturally define five kinds of hypervectors, as there are only five kinds of regular polyhedra in the three-dimensional space, namely, $\mathbf{V}=\{\mathbf{V}_4,\mathbf{V}_6,\mathbf{V}_8,\mathbf{V}_{12},\mathbf{V}_{20}\}$.
	
By using the definition of the hypervector above, we can easily classify the regular convex 4-polytope in terms of $\mathbf{V}_t^{[s]}$ as follows: (The index $s$ indicates the $s$-cell in the regular 4-polytope.)
\begin{align}
\mathbf{V}_t^{[s]}&=\left\{\mathbf{V}_t^{[5]},\mathbf{V}_t^{[8]},\mathbf{V}_t^{[16]},\mathbf{V}_t^{[24]},\mathbf{V}_t^{[120]},\mathbf{V}_t^{[600]}\right\}.
\end{align}
	
It is useful to note that, fortunately, each cell index $\ell$ is intimately concerned with the symbol $j$ in the regular convex 3-polytope. This observation offers the possibility for us to count the number of unitary sets for PQCs in secure quantum communication.

\bigskip	
\section{Relationship between qutrit-based PQCs and regular 4-polytope} \label{main}
In this section, we show that the qutrit-based PQCs can be related to the regular 4-polytope by using two strategies known from the Gell-Mann matrix ($\lambda_j$ with $j\in\{1,\ldots,8\}$) expansion and by applying quantum Fourier transform. The Gell-Mann matrix is a fundamental aspect of high-energy physics, and quantum Fourier transform is a core process in quantum algorithms. Although the Gell-Mann matrices are traceless and Hermitian in the $\tn{SU}(3)$ group, the modified Gell-Mann matrices have one exception in terms of the identity matrix (i.e., they are non-traceless). We also notice that one of the elements of the original Gell-Mann matrix (formally, $\lambda_8$) does not form a unitary matrix; however, our matrices all have unitary matrices. Therefore, our modified version could be more naturally considered as a generalization of the well-known Pauli matrix in $\tn{SU}(2)$.
	
\subsection{Generalized Gell-Mann matrices and qutrit-based PQCs}
	
As mentioned above, the PQCs over a single-qubit system are naturally constructed by a $2\times2$ unitary matrix ${R}_{\tn{c}}(\theta)\in\tn{SU}(2)$ in Eq.~\eqref{2d}, thus we can imagine a $3\times3$ unitary matrix in $\tn{SU}(3)$ for generating qutrit-based PQCs. To do this, we first consider the $3\times3$ orthogonal rotation matrix in $\tn{SO}(3)$ in the standard form of
\begin{widetext}
\begin{align} \label{3d}
R(\phi,\theta,\varphi)=\begin{pmatrix} \cos{\phi}\cos{\theta}\cos{\varphi}+\sin{\phi}\sin{\varphi} & -\cos{\phi}\cos{\theta}\sin{\varphi}+\sin{\phi}\cos{\varphi} & \sin{\theta}\cos{\phi} \\ -\sin{\phi}\cos{\theta}\cos{\varphi}+\cos{\phi}\sin{\varphi} & \sin{\phi}\cos{\theta}\sin{\varphi}+\cos{\phi}\cos{\varphi} & -\sin{\varphi}\sin{\theta} \\ -\sin{\theta}\cos{\varphi} & \sin{\theta}\sin{\varphi} & \cos{\theta} \end{pmatrix},
\end{align}
\end{widetext}
where each angle parameter is bounded by $0 \le \varphi < \pi$, $-\cfrac{\pi}{2}\le \theta \le \cfrac{\pi}{2}$, and $-\pi<\phi\le\pi$. Similar to the case of the two-dimensional complex rotation matrix ${R}_\tn{c}(\theta)$ in Eq.~\eqref{2d}, we can find a complex rotation matrix (or unitary matrix) $\check{R}_\tn{c}(\phi,\theta,\varphi)$ by extending the real rotation matrix in Eq.~\eqref{3d} with some complex numbers. However, this is not a simple task, and thus we have to change our intension of finding the complex rotation matrix to generalizing the well-known Gell-Mann matrices.
	
Here, we define a set of nine components in the $\tn{SU}(3)$ group, which can be modified from the original Gell-Mann matrix~\cite{G62}, and we call it the \emph{generalized} Gell-Mann matrix. All matrices are unitaries and Hermitian, and are explicitly given by
\begin{align} \label{gGM}
L_1&=\begin{pmatrix} 0 & 1 & 0 \\ 1 & 0 & 0 \\ 0 & 0 & 1 \end{pmatrix}, 
L_2=\begin{pmatrix} 0 & -i & 0 \\ i & 0 & 0 \\ 0 & 0 & 1 \end{pmatrix},
L_3=\begin{pmatrix} 0 & i & 0 \\ -i & 0 & 0 \\ 0 & 0 & 1 \end{pmatrix}, \nonumber\\
L_4&=\begin{pmatrix} 1 & 0 & 0 \\ 0 & 0 & 1 \\ 0 & 1 & 0 \end{pmatrix},
L_5=\begin{pmatrix} 1 & 0 & 0 \\ 0 & 0 & i \\ 0 & -i & 0 \end{pmatrix},
L_6=\begin{pmatrix} 1 & 0 & 0 \\ 0 & 0 & -i \\ 0 & i & 0 \end{pmatrix}, \\
L_7&=\begin{pmatrix} 0 & 0 & 1 \\ 0 & 1 & 0 \\ 1 & 0 & 0 \end{pmatrix},
L_8=\begin{pmatrix} 0 & 0 & i \\ 0 & 1 & 0 \\ -i & 0 & 0 \end{pmatrix},
L_9=\begin{pmatrix} 0 & 0 & -i \\ 0 & 1 & 0 \\ i & 0 & 0 \end{pmatrix}. \nonumber
\end{align}
	
By exploiting the generalized Gell-Mann matrix ($L$), we can construct an exact private quantum channel on the qutrit system. Formally, it is given by ($\forall \rho\in\cl{B}(\bb{C}^d)$)
\begin{equation} \label{ourPQC}
\Lambda_L(\rho)=\frac{1}{9}\sum_{j=1}^9L_j\rho L_j^\dag=\frac{\1}{3},
\end{equation}
where $L_j$'s are elements of the generalized Gell-Mann matrix. As mentioned above (see subsec.~\ref{definitions}), this construction is \emph{optimal} for the qutrit-based PQC~\cite{BR03,NK06,BZ07}. (Here, we notice that the matrix $L_j$ is equivalent to the key set component $k_j$.) The output of the channel $\Lambda_L$ is the exact maximally mixed state (i.e., $\frac{\1}{3}$), thus preserving its perfect secrecy. As an example, we can check that the resulting state of a qutrit, via $\Lambda_L$ in Eq.~\eqref{ourPQC}, is the maximally mixed state (see details in Appendix~\ref{app}). Thus, we conjecture that there exists another regular convex 4-polytope corresponding to the optimal PQC on the 3-level quantum system (see Table~\ref{table:pqc}).
	
Now, we need to consider the way in which to construct the PQCs for non-optimal cases (but preserving their security). We believe that other types of generalized Gell-Mann matrices exist that correspond to the PQCs of $\Lambda_S$, $\Lambda_H$, $\Lambda_{T_1}$, $\Lambda_O$, $\Lambda_D$, and $\Lambda_{T_2}$ with 20, 48, 96, 192, 1440, and 12 000-component basis vectors, respectively. (Note that the number of basis vectors is equivalent to the cardinality of the corresponding key set $K$.) To do this, we use the quantum Fourier transform strategy.

\subsection{Quantum Fourier transform and PQCs} \label{qutrit}
	
\begin{figure}[t!]
\includegraphics[width=\linewidth]{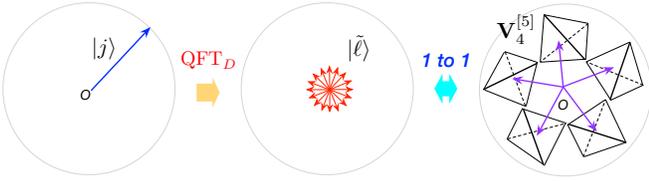}
\caption{One-to-one corresponding strategy as a result of extending the quantum Fourier transform to the hypervector. More precisely, a quantum state $\ket{j}$ can be transformed into a 20-component superposed state via $\tn{QFT}_{20}$, after which it is possible to match these to the hypervector $\mathbf{V}_4^{[5]}$ on the regular convex 4-polytope.}
\label{Fig2}
\end{figure}

The quantum Fourier transform (QFT), which is a linear transformation over quantum states, was first discovered by Coppersmith~\cite{C02}, and it was employed for many efficient calculations on quantum algorithms (e.g., Shor's algorithm~\cite{S97}). The Walsh–Hadamard or Hadamard transform is a special case of QFT on two-level quantum systems. Generally, the $d$-dimensional QFT, $\tn{QFT}_d$, is defined as follows: For any quantum state $\ket{j}\in\bb{C}^d$, the QFT is a map in the form of
\begin{equation}
\tn{QFT}_d:\ket{j}\mapsto\frac{1}{\sqrt{d}}\sum_{k=0}^{d-1}\omega^{jk}\ket{k},
\end{equation}
where $\omega=e^{\frac{2\pi i}{d}}$. In principle, it is possible to expand the sum of $k$ with $d$ elements to $\ell$ with $D(\ge d)$, and we call it an extended QFT in the $D$-dimension ($\tn{QFT}_D$), that is,
\begin{equation}
\tn{QFT}_D:\ket{j}\mapsto\frac{1}{\sqrt{D}}\sum_{\ell=0}^{D-1}\omega^{j\ell}\ket{\ell}\equiv|\tilde{\ell}\rangle.
\end{equation}
Here, we notice that the total probability is conserved as 1, and each quantum state has a uniform probability distribution $\frac{1}{D}$. This extension enables us to find hypervectors as mentioned above (see subsec.~\ref{hypervec}). If we choose $D=20$, then $\tn{QFT}_D$ performs the following action, that is, for any $j$
\begin{equation*}
\tn{QFT}_{20}:\ket{j}\mapsto\frac{1}{\sqrt{20}}\sum_{\ell=0}^{19}\omega^{j\ell}\ket{\ell},
\end{equation*}
where $\ket{\ell}$ forms a basis vector over the Hilbert space of the output. After this QFT, we can easily convert each basis vector (by using clustering) into a hypervector in Eq.~\eqref{hypervector}. We depict this situation in Fig.~\ref{Fig2}, and we define this PQC as $\Lambda_S$ (see also Table~\ref{table:pqc}). We notice that the reason why the one-to-one correspondence is possible originates from the condition of the isotropic measure in Subsec.~\ref{measure}.
	
Conceptually, this approach allows us to obtain further generalizations for $\Lambda_H$, $\Lambda_{T_1}$, $\Lambda_O$, $\Lambda_D$, and $\Lambda_{T_2}$ as well as for high-dimensional PQC cases on any $d$-dimensional quantum state (i.e., qudit). 
	
\section{Conclusions} \label{conclusion}
In this study, we constructed and analyzed a private quantum channel (PQC) involving a three-level quantum system (i.e., qutrit) to the maximally mixed state, by exploiting two methods: the generalized Gell-Mann matrix and the modified quantum Fourier transform (QFT). For these constructions, we newly defined the notion of a hypervector on the regular convex 4-polytope, and found nine components of the generalized Gell-Mann matrix $L$ in the optimal case. Furthermore, we provided an expansion technique on QFT for non-optimal qutrit-based PQCs with the new notion of the hypervector. Here, a total of seven kinds of PQCs were presented, and each of the PQCs satisfied the security condition, that is, it produced the maximal von Neumann entropy in terms of $\frac{\1}{3}$. In fact, we can conclude that the power of the isotropic measure induces our results. 
	
A few intriguing open questions still remain with regard to the private quantum channel itself or beyond. The first one is that the optimal case of the qutrit-based PQC exactly predicts the component of unitary operations; however, we do not know what it is in the class of the regular convex 4-polytope. The second question relates to the way in which we can apply our results to the PQCs to highlight the research on higher dimensional geometry or quantum databases. Finally, our work is expected to contribute to establishing contact with mathematicians who are well acquainted with the quantum information sciences.
	
\bigskip
\section{Acknowledgments}
This work was supported by the Basic Science Research Program through the National Research Foundation of Korea through a grant funded by the Ministry of Science and ICT (Grant No. NRF-2020M3E4A1077861) and the Ministry of Education (Grant No. NRF-2018R1D1A1B07047512).
	
\section{Appendix} \label{app}
We describe the optimal private quantum channel with nine key-sets $K$ in Eq.~(\ref{gGM}) of the qutrit system. As a representative and exact example, if we choose $\check{\rho}=\frac{1}{2}\begin{pmatrix}1 & 0 & 0 \\ 0 & 1 & 0 \\ 0 & 0 & 0\end{pmatrix}$, then we can easily calculate the output (i.e., the maximally mixed state $\frac{\1}{3}$) of the private quantum channel $\Lambda_{L}$ as follows:
\begin{widetext}
\begin{align*} \label{eq:pqc}
\Lambda_L(\rho)&=\frac{1}{9}\sum_{j=1}^{9}L_{j}\check{\rho} L_{j}^\dag \nonumber\\
&= \frac{1}{18}\Bigg[\begin{pmatrix} 1 & 0 & 0\\ 0 & 1 & 0 \\ 0 & 0 & 0\end{pmatrix}+\begin{pmatrix} 1 & 0 & 0\\ 0 & 1 & 0 \\ 0 & 0 & 0\end{pmatrix}+\begin{pmatrix} 1 & 0 & 0\\ 0 & 1 & 0 \\ 0 & 0 & 0\end{pmatrix}+\begin{pmatrix} 1 & 0 & 0\\ 0 & 0 & 0 \\ 0 & 0 & 1\end{pmatrix}+\begin{pmatrix} 1 & 0 & 0\\ 0 & 0 & 0 \\ 0 & 0 & 1\end{pmatrix}+\begin{pmatrix} 1 & 0 & 0\\ 0 & 0 & 0 \\ 0 & 0 & 1\end{pmatrix}+\begin{pmatrix} 0 & 0 & 0\\ 0 & 1 & 0 \\ 0 & 0 & 1\end{pmatrix}+\begin{pmatrix} 0 & 0 & 0\\ 0 & 1 & 0 \\ 0 & 0 & 1\end{pmatrix}+\begin{pmatrix} 0 & 0 & 0\\ 0 & 1 & 0 \\ 0 & 0 & 1\end{pmatrix}\Bigg] \\
&= \frac{1}{3}\begin{pmatrix} 1 & 0 & 0\\ 0 & 1 & 0 \\ 0 & 0 & 1\end{pmatrix}=\frac{\1_3}{3}.
\end{align*}
\end{widetext}
In addition, it is possible to straightforwardly calculate the output for another input state of the qutrit ${\rho}$. We omit the results.

\bigskip
%

\end{document}